\documentclass[nofootinbib, english]{article}
\usepackage[T1]{fontenc}
\usepackage[latin9]{inputenc}
\usepackage{geometry}
\usepackage{textcomp}
\usepackage{amsmath}
\usepackage{amsfonts}
\usepackage{stackrel}
\usepackage{graphicx}
\makeatletter

\providecommand{\tabularnewline}{\\}

\makeatother
\usepackage{babel}

\begin{document}

\title{The universal meaning of the quantum of action \thanks{The German original was published as \textquotedbl{}Universelle Bedeutung des Wirkungsquantums\textquotedbl{} in Proceedings of Tokyo Mathematico-Physical Society 8, 1915, 106-116.\newline\newline
Credits of the translation: Karla Pelogia - Philosophisch - Historische Fakultät, Universität Stuttgart Keplerstr. 17, 70174 Stuttgart, Deutschland, karlapelogia@gmail.com\newline\newline
Carlos Alexandre Brasil - São Carlos Institute of Physics (IFSC), University of São Paulo (USP), PO Box 369, 13560-970 São Carlos, SP, Brazil, carlosbrasil.physics@gmail.com\newline\newline
Acknowledgements: C. A. Brasil wishes to thank Thaís V. Trevisan (Instituto de Física 'Gleb Wataghin' IFGW / Universidade Estadual de Campinas UNICAMP and School of Physics and Astronomy / University of Minnesota) and Reginaldo de J. Napolitano (IFSC/USP) for reading and criticizing the preliminary versions of this paper, Nícolas Morazotti (IFSC/USP) for his precious help in LaTeX, Miled Y. H. Moussa (IFSC/USP) for his hospitality along 2016 and to Programa Nacional de Pós-Doutorado (PNPD) from Coordenação de Aperfeiçoamento de Pessoal de Ensino Superior (CAPES). K. Pelogia would like to thank to Carlos Brasil, for this oportunity, to Dr. Klaus Hentschel (Universität Stuttgart) for his time to help us  and to Christian Wilde, Cristal da Rocha (USP) , James Payne, Sheila Clark: thank you for read and re read this paper always looking for mistakes and helping us to improve our paper.\newline\newline
Conventions: The $a,b,c$ letters between brackets indicate the original author's notes. The numbers indicate the translators analysis. Both the bibliographic citations by the author and by the translators obey the EPJH format. The numbers in the margins indicate the starts of original pages translated.\newline}}

\author{by Jun Ishiwara\thanks{deceased (1881-1947)}}
\date{}
\maketitle
\begin{center}
  [Submitted on April 4th, 1915.]
\end{center}
Since the theory of radiation demanded the existence of a universal 
action quantum $h$, different developmental phases have taken place.
It would currently be most appropriate to select the following point of
view of a universal meaning of $h$: to explain the occurrence of
$h$ by the existence of certain finite elementary regions of the
state space \textbf{}\footnote{\samepage \textbf{ }Ishiwara uses \textquotedbl{}state space\textquotedbl{}
instead \textquotedbl{}phase space\textquotedbl{} - we will discuss
this detail in section III of \cite{Pelogia}.}, whose size for all different elementary regions of equal probability
is $h$ \cite{Planck3,Sommerfeld6}\footnote{\textbf{}Sommerfeld's, \textquotedbl{}The Planck's constant
and its overall importance for the molecular physics\textquotedbl{},
was presented on 25 September 1911 at the \textquotedbl{}83rd Scientific
meeting in Karlsruhe\textquotedbl{}, published in \emph{Physikalische
Zeitschrift }\cite{Sommerfeld6}, a German magazine published from
1899 to 1945. In the book of Planck, \textquotedbl{}Lectures on the
theory of heat radiation\textquotedbl{} \cite{Planck3}, Ishiwara
specifies chapter 3 \textquotedbl{}Entropy and probability\textquotedbl{}. }. However, the constant $h$ relates not only to stationary \marginpar{107}radiation
processes, but also must be regarded as manifestation of a universal molecular
property. Of great interest in this context is the Bohr atomic model,
where to the quantity $\frac{h}{2\pi}$ is attributed a new meaning
as the quantum of angular momentum of the rotating electrons \cite{Bohr1,Bohr2,Bohr3,Bohr4}
\textbf{}\footnote{\textbf{ }Ishiwara refers not only to the classic set of papers \cite{Bohr1,Bohr2,Bohr3}
where Bohr presents his atomic model, but also to a paper \cite{Bohr4}
where he tries to explain the Stark and Zeeman effects and the spectral
lines of other elements - it is worth noting that Bohr, in this work,
had proposed the possibility of elliptical orbits for the electron,
but did not consider it explicitly in his calculations. See 
section III of \cite{Pelogia} for more details about it.}. However, there has been as yet no theoretical attempt to identify a possible
link between the two mentioned meanings of the same constant. Also,
arisen quite independently, is the Sommerfeld hypothesis that the
action integral occurring in the Hamiltonian principle is equal to
$\frac{h}{2\pi}$ for every pure molecular process \cite{Debye1,Sommerfeld6}.

I would like to propose the following question: Are those different interpretations
of $h$ identical to each other? If not, it should be asked: Which
assessment of the universal meaning of $h$ is the right one? Should
all phenomena be explained by one and the same basic assumption? 

This paper will hopefully provide some answers.

\section*{1. The basic assumption }

Bohr's hypothesis always equates the angular momentum of the rotating
electrons in the atom with an entire multiple of the quantity $\frac{h}{2\pi}$
. Even if this is correct, however, it should not be understood as
the actual meaning of $h$, because the angular momentum is a vector
quantity whose absolute value can generally not simply be put together
from elementary parts linearly. A more fundamental definition of the
scalar quantity $h$ will unfold differently in any event. 

From the same perspective the Sommerfeld Rule seems generally valid
because it identifies $h$ with the invariant of mechanics that is
the action size (up to the factor of $\frac{h}{2\pi}$ ). Sommerfeld
has used it to explain the photoelectric effect; an application thereof
to the statistics of the radiation unfortunately will not produce
the desired result.
\footnote{ \textbf{} In the 1911 paper,
Sommerfeld emphasizes the use of the expression ``quanta of action'', instead of
``quanta of energy''. Here, he compares the classical (with the Rayleigh's formula) and
quantum (with the Planck's formula) for black-body theory and shows
the limits where both expressions coincide. After he presents his quantization
hypothesis 

\begin{equation*}
\int_{0}^{\tau}L\,dt=\frac{h}{2\pi}\label{Sommerfeld} 
\end{equation*}
and analyses the photoelectric effect, Sommerfeld tries to expand it
to relativistic systems. At the end, there is a discussion with Johannes
Stark (1874-1957), Einstein, Leo K?nigsberger (1837-1921) and Heinrich
Rubens (1865-1922). The second paper cited here, the work with his
student Peter Debye (1884-1966), \textquotedbl{}Theory of the photoelectric
effect from the standpoint of the quantum of action\textquotedbl{},
is a more extensive and detailed version of the theory for photoelectric
effect exposed on the first one. The theory of Debye-Sommerfeld for
photoelectric effect was not equivalent to the Einstein's one - see
section III of [our analysis paper] for comments about this
point.}

After futile attempts now only one path remains along which for me
to proceed. I formulate the following basic assumption: \marginpar{108}

\textquotedbl{}An elementary structure of the matter or a system of
immense number of elementary structures exist in a stationary periodic
motion or in a statistical equilibrium. The state is completely determined
by coordinates $p_{1}$, $p_{2}$,..., $p_{f}$ and the associated
\emph{momentum} coordinates $q_{1}$, $q_{2}$, ..., $q_{f}$ \footnote{\textbf{ }Ishiwara writes against the popular convention, where the
$q$-letter indicates the position and the $p$-letter indicates the
\emph{momentum}.}. Then the movements in nature will always take place in such a way
that a decomposition of each state plane $p_{i}q_{i}$ in those elementary
areas of equal probability is allowed, whose mean value in a certain
point of the state space:

\begin{equation}
h=\frac{1}{j}\underset{i=1}{\overset{j}{\sum}}\int q_{i}\,dp_{i}\label{eq1}
\end{equation}
equals a universal constant.\textquotedblright{} \textbf{}\footnote{\samepage\textbf{ }The condition of Ishiwara is different from the condition
proposed by Bohr \cite{Waerden}, Sommerfeld \cite{Sommerfeld3,Sommerfeld4,Sommerfeld5}
and Wilson \cite{Wilson},
\begin{equation*}
\int q_{i}\,dp_{i}=n_{i}\,h\label{aux1} 
\end{equation*}
where, for $j$ degrees of freedom, we have $j$ equations, with $i$
varying from $1$ to $j$ - i. e., there are no more summation over
$i$, and no division by $j$ \cite{Abiko}. In (\ref{aux1}), we
follow the convention where the $q$-letter indicates the position
and the $p$-letter indicates the \emph{momentum}. See a detailed
discussion in section III of \cite{Pelogia}.}

The application of the last sentence to the statistics of black-body
radiation on the one hand and the gas of molecules on the other hand
already provided an experimental proof of the universality of $h$.
The value of $h$ calculated by O. Sackur \cite{Sackur1,Sackur2}
and H. Tetrode \cite{Tetrode1,Tetrode2} from the chemical constants of argon and mercury vapor is in excellent
agreement with the value that is well-known from radiation \textbf{}\footnote{\textbf{ }In this paragraph, Ishiwara refers to two papers of Otto
Sackur (1880-1914), \textquotedbl{}The importance of the elementary
quantum of action for the theory of gases and the calculation of chemical
constants\textquotedbl{}, that was published in the book \textquotedbl{}In
Honour W. Nernst to his 25-year-old Philosophical Doctoral Jubilee\textquotedbl{},
and \textquotedbl{}The universal significance of the so-called 'elementary
quantum of action' \textquotedbl{}; and to two papers of Hugo Tetrode
(1895-1931), \textquotedbl{}The chemical constant of the gases and
the elementary quantum of action\textquotedbl{} and \textquotedbl{}Correction
to my work : ' The chemical constant of the gases and the elementary
quantum of action '\textquotedbl{}. They talk about the \emph{chemical
constant,} a constant which occurred in the expression for the absolute
entropy of a gaseous substance, that was used by them to the quantization
of translational motion of atomic particles on monoatomic gases and
to analyze the deviation of the classical equation at low temperatures.
\cite{Mehra2,Mehra} At this time, the expression 'value of radiation' would
have referred not necessarily to $h$ itself, but to other constants used
at that time and that could be related to Planck's one. For example,
in his original work about photoelectric effect \cite{Einstein4},
Einstein did not used $h$, but $R$ (the universal gas constant),
$\alpha$ and $\beta$ (the constants that appeared on the original
expression of Planck for the black-body spectra \cite{Feldens,Planck1,Planck2,Studart}).
The papers of Sackur and Tetrode, cited by Ishiwara, are illustrative
about it. On the other side, Robert Andrew Millikan (1868-1953), in his
studies about the photoelectric effect, was making measurements of
$h$ itself - and his results were cited by Sommerfeld \cite{Debye1,Sommerfeld3}
in the name of Millikan. Sometimes, the name cited is James Remus
Wright (1883? - 1937), a member of Millikan's team who obtained his
Ph. D. in 1911 (his thesis \textquotedbl{}The positive potential of
aluminum as a function of the wave-length of the incident light\textquotedbl{}
\cite{Wright1} is only his paper published in Physical Review \cite{Wright2})
and, after that he held the position of Chief of Department at the University of Philippines
from 1911 to 1919, working with new oil refining processes at the Standard
Oil Company from 1919 until two years before his death \cite{AMS,Wright3}.}.

It should now be examined whether the statement is really applicable
also to the process of individual electron movements, especially to
the case of the Bohr atom model and the photoelectric problem. Before
I concentrate on each part individually, I would like to outline a
general observation. 

Imagine an electron; its position being expressed by the radial vector
$r$ that is drawn from the point of origin. By marking the differentiation
with respect to the (Minkowski) proper time $\tau$ \textbf{}\footnote{\textbf{}Here, the reference to \textquotedbl{}proper time\textquotedbl{}
is significant in the concerns of Ishiwara with relativity\cite{Einstein1,Hu}.
Indeed, Ishiwara dedicated most of his time to Einstein's relativity,
and he was a scholar and commentator on the theory, gaining reputation
both in Japan and China, where he was known as \textquotedblleft \emph{the
only expert of relativity studies in Japan} (\dots )\emph{ }{[}who{]}\emph{
}had a unique understanding of the theory of relativity\textquotedblright .
The first paper by Ishiwara on the theory was published in 1909 and
was the first in Japan on this topic. \cite{Hu,Sigeko} }by a top marked point, the momentum quantity of the electron is given
by $m_{0}\dot{r}$ with $m_{0}$ denoting the rest mass. The quantum
theorem stipulates here that the integral

\begin{equation}
h=\frac{1}{j}\int\left(m_{0}\dot{r},\,dr\right)\label{eq1linha}\tag{1'}
\end{equation}
\marginpar{109}has a universal value in an elementary region with
equal probabilities. Assuming $j$ equals to 1, 2 or 3, $j_{\theta}$
depending on the movement being one-, two- or three-dimensional. 

Now, the area of equal probability is likely to be bounded by two
specific stationary movements. By extending the integral (\ref{eq1linha})
over the whole area, which is bounded by one of said stationary movements,
it must equal an integer multiple of $h$; thus

\begin{equation}
n\,j\,h=\int m_{0}\dot{r}\,dr\label{eq2}
\end{equation}
when $n$ is an integer. By looking at $r$ as a function of the proper
time $\tau$, one can rewrite the integral: 

\[
\int m_{0}\dot{r},\,dr=\int m_{0}\dot{r}^{2}\,d\tau
\]
By putting \textbf{}\footnote{\textbf{}In the equation (\ref{eq3}) there is a typographical
error, appearing $r^{2}$ instead $\dot{r}^{2}$, but it causes no
effects in further developments, because the correct expression for
kinetic energy is used. }

\begin{equation}
T=\frac{1}{2}m_{0}r^{2}\label{eq3}
\end{equation}
(\ref{eq2}) becomes in 

\begin{equation}
n\,j\,h=2\int_{0}^{\theta}T\,d\tau\label{eq4}
\end{equation}
where $\theta$ is the period of the considered movement. 

It is further more

\[
\int_{0}^{\theta}m_{0}\dot{r}^{2}\,d\tau=\left|m_{0}r\,\dot{r}\right|_{0}^{\theta}-\int_{0}^{\theta}m_{0}\left(r\,\ddot{r}\right)\,d\tau
\]
The first term on the right vanishes because of the periodicity; however,
the second is equal to twice the time integral of the Clausius Virial
$V$ because 

\begin{equation}
m_{0}\left(\ddot{r}\,r\right)=\left(\mathfrak K r\right)=-2V\label{eq5}
\end{equation}
is to be set. In the central force whose magnitude is inversely proportional
to the square of the distance $V$ equals the potential of the force
$\Re$. Therefore relation (\ref{eq4}) may also be written as:

\begin{equation}
n\,j\,h=2\int_{0}^{\theta}\left(T+V\right)\,d\tau\label{eq6}
\end{equation}

From the above it is clear that the classic action principle:

\[
\delta\int\left(T-V\right)\,d\tau=\delta\int m_{0}c^{2}d\tau=0
\]
\marginpar{110}still remains valid here, independent of the quantum
theorem. Therefore it seems to me that the Sommerfeld hypothesis is
losing its basic principle.

\section*{2. The Bohr model of the atom}

The following text will explain the application of the quantum theorem
to stationary motion of electrons inside the atom. For the sake of
simplicity we focus on the atomic nucleus and an orbiting electron
and leave the effect of the other electrons aside.

\newcounter{tempfootnote}
\setcounter{tempfootnote}{\value{footnote}}
\setcounter{footnote}{0}
\renewcommand{\thefootnote}{\alph{footnote}}
\makeatletter%
\long\def\@makefnmark{%
\hbox {{\normalfont \@thefnmark }}}%
\makeatother

Following Coulomb's law the electron carries out the central movement
around the nucleus of $O$ (\footnote{ \textbf{} I neglect the terms proportional to the square of the velocity
  against that of the speed of light $c$.}); 

\setcounter{footnote}{\value{tempfootnote}}
\renewcommand{\thefootnote}{\arabic{footnote}}
\makeatletter%
\long\def\@makefnmark{%
\hbox {{\textsuperscript \@thefnmark }}}%
\makeatother

so it describes as a focal point an ellipse with greater semiaxis
$a$ and the eccentricity $\varepsilon$. With $O$ being the point
of origin and the elliptical plane being the $xy$ plane, the path
is expressed by equation\footnotemark \textbf{}

\footnotetext{\textbf{}The equation (\ref{eq7}) describes an ellipse with his
  left focus ($F=\varepsilon a$) on the origin $O$ of the system of
  coordinates, with greater semiaxis over the $x$-axis, as indicated
  on the Fig. 1.
  
  \begin{tabular}{l}
    \includegraphics[scale=0.5]{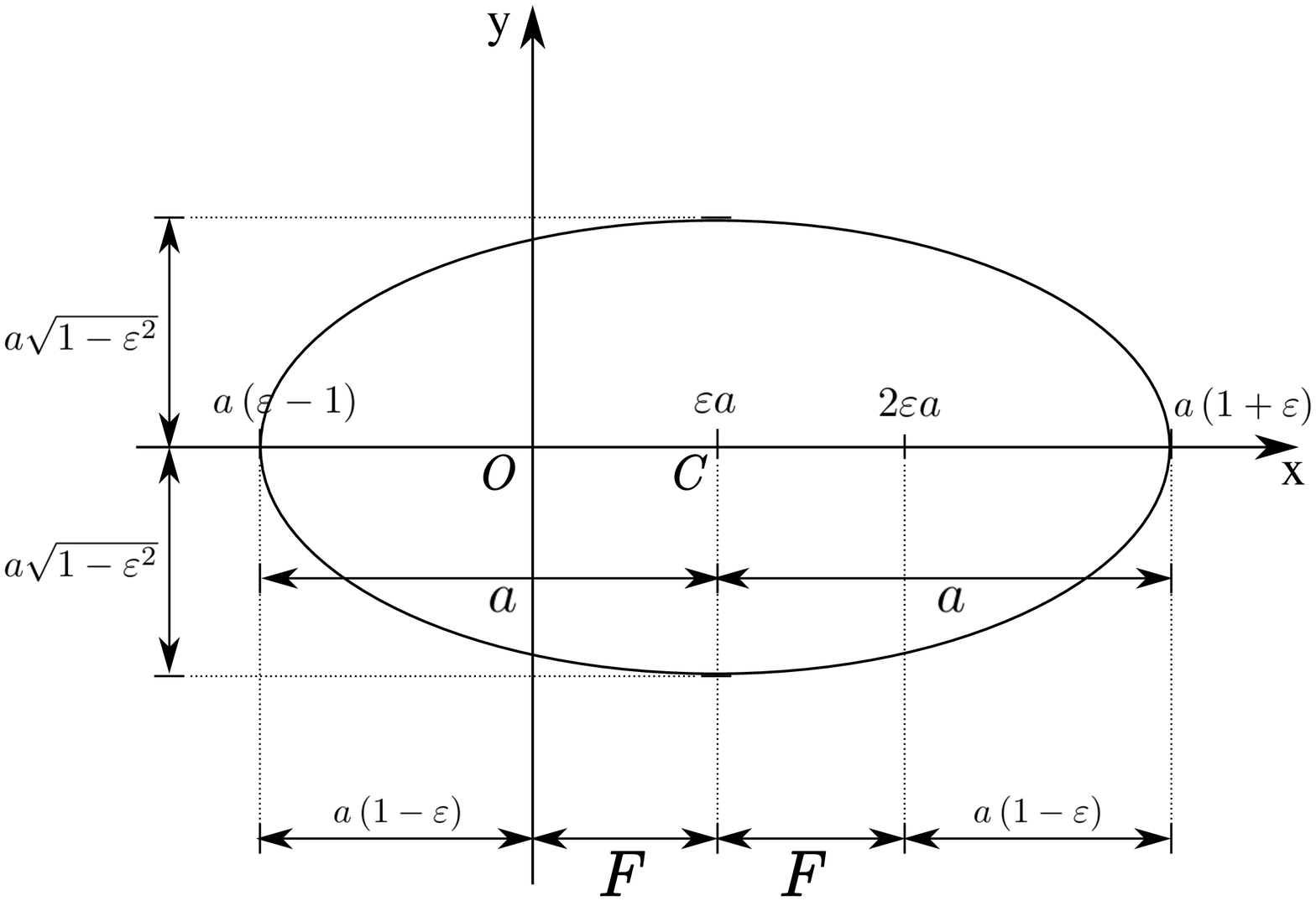}\tabularnewline
    Figure 1: Ellipse described by the equation (\ref{eq7}), with center
    on the $C$ point and origin $O$ of the system of coordinates over
    the left focus.\tabularnewline
  \end{tabular}
}

\begin{equation}
\left(x+\varepsilon a\right)^{2}+\frac{y^{2}}{1-\varepsilon^{2}}=a^{2}\label{eq7}
\end{equation}
The law of areas then can be expressed as

\begin{equation}
m_{0}\left(x\dot{y}-y\dot{x}\right)=f\label{eq8}
\end{equation}
$f$ means the constant angular momentum. 

Differentiating (\ref{eq7}) with respect to the proper time, we obtain

\begin{equation}
\left(x+\varepsilon a\right)\dot{x}+\frac{y}{1-\varepsilon^{2}}\dot{y}=0\label{eq9}
\end{equation}
Elimination of $y$ and $\dot{y}$. from (\ref{eq7}), (\ref{eq8})
and (\ref{eq9}) yields

\[
m_{0}^{2}\left(1-\varepsilon^{2}\right)\left\{ a^{2}\left(1-\varepsilon^{2}\right)-\varepsilon ax\right\} ^{2}\dot{x}^{2}=f^{2}\left\{ a^{2}-\left(x+\varepsilon a\right)^{2}\right\} 
\]
which through the introduction of the variable \textbf{}\footnote{\textbf{ }In the equation defining the new variable there is a typographic
error, the expression truly used by Ishiwara is $x'=x+\varepsilon a$.
This change of variable is equivalent to set the origin $O$ of the
system of coordinates on the center $C$ of the ellipse.}

\[
x'=x+a
\]
can be rewritten as:
\[
m_{0}^{2}\left(1-\varepsilon^{2}\right)a^{2}\left(a-\varepsilon x'\right)^{2}\dot{x}'^{2}=f^{2}\left(a^{2}-x'^{2}\right)
\]
After taking the square root

\begin{equation}
m_{0}\dot{x}'=\mp\frac{f\sqrt{a^{2}-x'^{2}}}{a\sqrt{1-\varepsilon^{2}}\left(a-\varepsilon x'\right)}\label{eq10}
\end{equation}

By eliminating $\dot{x}$ one easily finds \textbf{}\footnote{\textbf{ }To arrive at (\ref{eq11}), we must find the expression
for $\dot{y}$ 
\begin{equation*}
\dot{y}^{2}=\frac{\left(1-\varepsilon^{2}\right)\left(x+\varepsilon a\right)^{2}}{a^{2}-\left(x+\varepsilon a\right)^{2}}\dot{x}^{2} 
\end{equation*}
}:

\marginpar{111}
\begin{equation}
m_{0}\dot{y}=\frac{fx'}{a\left(a-\varepsilon x'\right)}\label{eq11}
\end{equation}

The formula (\ref{eq2}) thus writes itself, since $j=2$ to set,

\[
2nh=\int m_{0}\dot{x}'\,dx'+\int m_{0}\dot{y}\,dy
\]
or because of the symmetry of the figures with respect to the $x'$-axis

\[
nh=\int_{-a}^{a}m_{0}\dot{x}'\,dx'+\int_{-a}^{a}m_{0}\dot{y}\,\frac{dy}{dx'}dx'
\]
These integrals are calculated according to (\ref{eq7}), (\ref{eq10})
and (\ref{eq11}) as follows \textbf{}\footnote{\textbf{ }There is another typographical error on the following
equation, instead $\frac{\left(1-\varepsilon^{2}\right)x'}{\left(a-\varepsilon x'\right)\sqrt{a^{2}-x'^{2}}}$,
the second integrand is $\frac{\left(1-\varepsilon^{2}\right)x'^{2}}{\left(a-\varepsilon x'\right)\sqrt{a^{2}-x'^{2}}}$.}:

\begin{eqnarray*}
nh & = & \frac{f}{a\sqrt{1-\varepsilon^{2}}}\int_{-a}^{a}\left\{ -\frac{\sqrt{a^{2}-x'^{2}}}{a-\varepsilon x'}-\frac{\left(1-\varepsilon^{2}\right)x'}{\left(a-\varepsilon x'\right)\sqrt{a^{2}-x'^{2}}}\right\} dx'\\
 & = & -\frac{f}{a\sqrt{1-\varepsilon^{2}}}\int_{-a}^{a}\frac{a+\varepsilon x'}{\sqrt{a^{2}-x'^{2}}}dx'=\frac{\pi f}{\sqrt{1-\varepsilon^{2}}}
\end{eqnarray*}
Thus we conclude

\begin{equation}
f=n\frac{h}{\pi}\sqrt{1-\varepsilon^{2}}\label{eq12}
\end{equation}
The angular momentum of the electron motion therefore equals an integral
multiple of $\frac{h}{\pi}\sqrt{1-\varepsilon^{2}}$. In the case
where the elliptical orbit turns into a circle, it obtains only 

\begin{equation}
f=n\frac{h}{\pi}\label{eq12linha}\tag{12'}
\end{equation}
Thus it is shown not only that the Bohr hypothesis generally fails,
but also that it still is not quite correct even for the circular
motion. We will now examine how this modification of the basic assumption has further consequences. 

The period of the electron motion is given by \textbf{}\footnote{\textbf{ }The period $\theta$ of the movement by a central conservative
force is given by \cite{Fetter} 
\begin{equation}
\theta=2\pi\frac{m_{0}ab}{f} \nonumber
\end{equation}
where $b=a\sqrt{1-\varepsilon^{2}}$ is the minor semi-axis of the
ellipsis, then we have in a direct way the result. However, Ishiwara
adopts the time integration, with $\tau=\tau\left(x',y\right)$. It
is easy to see that $d\tau=2\frac{dx'}{\dot{x}'}$ and, with (\ref{eq10}),
we can obtain the period. However, there is a typographical error
in the intermediate expression, the correct form is
\begin{equation*}
\theta=\int d\tau=-\frac{2a\,m_{0}\sqrt{1-\varepsilon^{2}}}{f}\int_{-a}^{a}\frac{a-\varepsilon x\text{\textasciiacute}}{\sqrt{a^{2}-x'^{2}}}dx\text{\textasciiacute}
\end{equation*}
}

\begin{eqnarray*}
\theta & = & \int d\tau=-\frac{4a\,m_{0}\sqrt{1-\varepsilon^{2}}}{f}\int_{-a}^{a}\frac{a-\varepsilon x\text{\textasciiacute}}{\sqrt{a^{2}-x'^{2}}}dx\text{\textasciiacute}\\
 & = & \frac{2\pi a^{2}m_{0}\sqrt{1-\varepsilon^{2}}}{f}
\end{eqnarray*}

Designating the number of vibrations per unit of time with $\nu$,
it follows thereafter 

\marginpar{112}
\begin{equation}
\frac{f}{\sqrt{1-\varepsilon^{2}}}=2\pi a^{2}m_{0}\nu\label{eq13}
\end{equation}
and with (\ref{eq12}) 

\begin{equation}
\frac{n\,h}{\pi}=2\pi a^{2}m_{0}\nu\label{eq14}
\end{equation}

On the other hand, if the electric charge of the revolving electron
is $-e$ and for the atomic nucleus is $e'$, we have, as is well-known
\textbf{}\footnote{\textbf{ }The equation (\ref{eq15}) can be obtained considering the conservation of energy under
Coulombian potential.},

\begin{equation}
a\left(1-\varepsilon^{2}\right)=\frac{f^{2}}{m_{0}e\,e\text{\textasciiacute}}\label{eq15}
\end{equation}
From (\ref{eq12}) and (\ref{eq15}) it follows that

\begin{equation}
a=\frac{n^{2}h^{2}}{\pi^{2}m_{0}e\,e\text{\textasciiacute}}\label{eq16}
\end{equation}
Finally, by eliminating $a$ and solving for $\nu$ one obtains from
(\ref{eq14}) and (\ref{eq16}) 

\begin{equation}
\nu=\frac{\pi^{2}m_{0}e^{2}\,e\text{\textasciiacute}^{2}}{2n^{3}h^{3}}\label{eq17}
\end{equation}

The average kinetic energy of the system according to (\ref{eq4})
with the value $j=2$, amounts to:

\begin{equation}
\overline{T}=n\,h\,\nu\label{eq18}
\end{equation}
or by inserting $\nu$ from (\ref{eq17}) 

\begin{equation}
\overline{T}=\frac{\pi^{2}m_{0}e^{2}\,e\text{\textasciiacute}^{2}}{2n^{2}h^{2}}\label{eq19}
\end{equation}

Assuming different integers for $n$, one gets a series of values
that correspond to states of different probabilities. Yet only during
the transition between two such states can the electron emit a radiation
energy \footnote{\textbf{ }According to the atomic model of Bohr. \cite{Bohr1,Bohr2,Bohr3,Calouste,Parente}}.
The vibration frequency $\nu_{1}$ of the emitted radiation is to
be determined herein so that the energy output is exactly equal to
$h\nu_{1}$. According to (\ref{eq19}) \textbf{}\footnote{\textbf{ }We can see here that the equation (\ref{eq20}), if applied
to Hydrogen atom ($e\text{\textasciiacute}=e$), will furnish a frequency
of transition
\begin{equation*}
\nu_{1}=\frac{\pi^{2}m_{0}e^{4}}{2h^{3}}\left(\frac{1}{n_{1}^{2}}-\frac{1}{n_{2}^{2}}\right)
\end{equation*}
different from the real one - compare with equation (4) of \cite{Bohr1}
and (\ref{eq21})
\begin{equation}
\nu_{1}=\frac{2\pi^{2}m_{0}e^{4}}{h^{3}}\left(\frac{1}{n_{1}^{2}}-\frac{1}{n_{2}^{2}}\right) \nonumber
\end{equation}
which was obtained by Sommerfeld with his formula (\ref{aux1}) (see
footnote 5).}

\begin{equation}
\nu_{1}=\frac{\pi^{2}m_{0}e^{2}\,e\text{\textasciiacute}^{2}}{2h^{3}}\left(\frac{1}{n_{1}^{2}}-\frac{1}{n_{2}^{2}}\right)\label{eq20}
\end{equation}

To identify this law with Balmer\textquoteright s theory, one now
has to assume that the central charge of the hydrogen atom consists
of two electrical elementary quanta. For if one sets \textbf{}\footnote{\textbf{ }Obviously, this is a hypothesis contrary to the very definition
of a chemical element and it is not necessary if we use the Bohr-Sommerfeld-Wilson
version, equation (\ref{aux1}). For a more detailed analysis of the original
Sommerfeld calculation, see \cite{Castro}.}, 

\[
e\text{\textasciiacute}=2e
\]
\marginpar{113}then (\ref{eq20}) turns into

\begin{equation}
\nu_{1}=\frac{2\pi^{2}m_{0}e^{4}}{h^{3}}\left(\frac{1}{n_{1}^{2}}-\frac{1}{n_{2}^{2}}\right)\label{eq21}
\end{equation}
which is just what is found in Bohr\textquoteright s theory. 

The fact that the hydrogen atom in actual fact in the neutral state contains
two electrons is in my opinion still significantly supported by the
following circumstances:
\begin{enumerate}
\item The adoption of two rotating electrons can completely explain the
large value of the magnetic susceptibility of liquid hydrogen according
to the Langevin theory of diamagnetism. \footnote{\textbf{ } A contemporary reference about the diamagnetism is \cite{Essen} and, for Langevin's theory in particular, see the profound analysis of Navarro and Olivella \cite{Navarro}. }
\item If one takes addictional quantities of the order of $\frac{1}{c^{2}}\left(\frac{dr}{dt}\right)^{2}$
into consideration, one obtains a more rigorous formula for $\nu_{1}$,
which is similar to that derived by Allen \cite{Allen1}, and fits
the precise spectral measurements better \cite{Allen2}.\textbf{}\footnote{\textbf{ }Several points can be analyzed here. The relativistic
treatment, that Ishiwara says he wants to share in a forthcoming work,
was conducted soon after the publication of this paper, by Sommerfeld
\cite{Sommerfeld4,Sommerfeld2}, and allowed to explain the fine structure
of spectral lines. Ishiwara cites the works of Herbert Stanley Allen
(1873 - 1954), who suggests, to obtain more precise spectral formulas,
following on from the previous work of other researchers, to treat the atom as
consisting of a \emph{magnetic} core, electrically charged and surrounded
by one or more electrons, with the quantization of the electronic
orbital angular momentum. In the first paper, however, the conclusion
is, \textquotedbl{}that the magnetic forces set up by the atom are
not in themselves sufficient to account for more than a small fraction
of the effect that would be necessary to give the observed distribution
of lines in spectral series\textquotedbl{} \cite{Allen1}. In the
second paper, published in the same volume as the first one, Allen
supposes that the magnetic moment of the nucleus has different values
for the steady states of motion but, one more time, the model just
\textquotedbl{}appears that it is possible to account for the series
spectrum of hydrogen\textquotedbl{} \cite{Allen2}.}
\end{enumerate}
The numerical figures regarding this I intend to communicate in a forthcoming
work.

\section*{3. The photoelectric phenomenon}

Our fundamental law also seems to show its special power in regards to the explanation of photoelectric phenomenon. We first imagine the photoelectric
process after Sommerfeld \cite{Debye1} \textbf{}\footnote{\textbf{ }Once more, is important to note that Ishiwara follows
the theory of the photoelectric effect by Sommerfeld, not Einstein's
one. See section III of \cite{Pelogia}. }, as follows: 

An electron is resonant with incident radiation, so a large amount
of the energy is accumulated in it, but only until it reaches a certain
value, which could then satisfy the equation (\ref{eq4}), if the
incident radiation would stop at just the same moment and the electron
would still continue its stationary motion inside the atom.

\setcounter{tempfootnote}{\value{footnote}}
\setcounter{footnote}{1}
\renewcommand{\thefootnote}{\alph{footnote}}
\makeatletter%
\long\def\@makefnmark{%
\hbox {{\normalfont \@thefnmark }}}%
\makeatother

If one imagines the simplest case of perfect resonance, one can write
for the linear oscillation of the electron along the $x$-axis (\footnote{\textbf{ }Neglecting the damping.}):
\setcounter{footnote}{\value{tempfootnote}}
\renewcommand{\thefootnote}{\arabic{footnote}}
\makeatletter%
\long\def\@makefnmark{%
\hbox {{\textsuperscript \@thefnmark }}}%
\makeatother

\begin{equation}
\ddot{x}+4\pi^{2}\nu^{2}x=\frac{eC}{m_{0}}\cos\left(2\pi\nu\tau\right)\label{eq22}
\end{equation}
\marginpar{114}where $C$ is the amplitude of the incident electric
field, $\nu$ the number of vibrations per unit of time. 

With the initial conditions: 

\[
x=0\;\mathrm{and}\;\dot{x}=0\;\mathrm{f}\mathrm{or}\;t=0
\]
one gets the solution \textbf{}\footnote{\textbf{ }The simple form of the solution (\ref{eq23}) is due to:
1) the absence of the damping term on (\ref{eq22}), as emphasized
in the footnote {[}b{]}, 2) the form of initial conditions and 3)
the fact that the frequency of the incident wave is the same as the
natural frequency of the electron (resonance case).}: 

\begin{equation}
x=\frac{eC}{4\pi m_{0}\nu}\tau\sin\left(2\pi\nu\tau\right)\label{eq23}
\end{equation}
and thus the kinetic and the potential energy of the oscillating electron
results in: 

\begin{equation}
T=\frac{1}{2}m_{0}\dot{x}^{2}=\frac{e^{2}C^{2}}{32\pi^{2}m_{0}\nu^{2}}\left[\sin\left(2\pi\nu\tau\right)+2\pi\nu\tau\cos\left(2\pi\nu\tau\right)\right]^{2}\label{eq24}
\end{equation}
\begin{equation}
U=2\pi^{2}m_{0}\nu^{2}x^{2}=\frac{e^{2}C^{2}}{8m_{0}}\left[\tau\sin\left(2\pi\nu\tau\right)\right]^{2}\label{eq25}
\end{equation}

The electron now separates itself from the atom after an accumulation
time $\theta_{1}$ has elapsed. Now, the time $\theta_{1}$ is determined
according to our hypothesis as follows: 

We assume in our thoughts that the electron, from the time $\tau=\theta_{1}$
onwards continues with stationary vibrations. Then the quantum law
(\ref{eq4}) should apply. Since $j=1$ in our example, it reads \textbf{}\footnote{\textbf{ }Here, as Ishiwara considers $j=1$, there will be no influences
of his incorrect condition (\ref{eq1}) - see footnote 5. }

\begin{equation}
nh=2\int_{0}^{\theta}T\,d\tau\label{eq26}
\end{equation}
where $\theta$ means the period of the considered vibrations. This
means instead of (\ref{eq22}) we have

\[
\ddot{x}+4\pi^{2}\nu^{2}x=0
\]
which results in:

\[
x=a\sin\left(2\pi\nu\tau+\delta\right)
\]
It therefore is calculated:

\[
T=\frac{1}{2}m_{0}\dot{x}^{2}=2\pi^{2}m_{0}\nu^{2}a^{2}\cos^{2}\left(2\pi\nu\tau+\delta\right)
\]
and therefore 

\[
\int_{0}^{\theta}T\,d\tau=\pi^{2}m_{0}\nu a^{2}
\]
With this value one obtains from (\ref{eq26})

\marginpar{115}
\begin{equation}
a^{2}=\frac{nh}{2\pi^{2}m_{0}\nu}\label{eq27}
\end{equation}
the same relationship as in (\ref{eq14}).

On the other side, the amplitude $a$ is the one which the resonant oscillation (\ref{eq23}) assumes at the time $\tau=\theta_{1}$. Hence
\textbf{}\footnote{\textbf{ }If $x\left(\theta_{1}\right)$ is the largest value of
(\ref{eq23}), we must have $\sin\left(2\pi\nu\theta_{1}\right)=1$.}

\begin{equation}
a=\frac{eC}{4\pi m_{0}\nu}\theta_{1}\label{eq28}
\end{equation}
Comparing (\ref{eq27}) and (\ref{eq28}), the result is 

\begin{equation}
\theta_{1}=\frac{\sqrt{8m_{0}\nu nh}}{eC}\label{eq29}
\end{equation}
It is remarkable that this value of the accumulation time completely
agrees with the value calculated from the Sommerfeld theory, although
the basic assumption of the latter theory is different from ours.
\textbf{}\footnote{\textbf{ }Here, where Ishiwara talks about Sommerfeld's theory, he
is talking about the Eq. (\ref{Sommerfeld}), presented by the first
time at the first Solvay Congress \cite{Jammer,Straumann}, not to
the ``definitive'' version, 
\begin{equation*}
\int q_{i}\,dp_{i}=n_{i}\,h\label{aux2}
\end{equation*}
}

The difference between the two theories shows, of course, in other relationships,
i.e., if one asks for the energy of the electron at the end of the
accumulation time. At time $\tau=\theta_{1}$ the kinetic energy is
according to (\ref{eq24}) $T=\frac{e^{2}C^{2}}{32\pi^{2}m_{0}\nu^{2}}\left[\sin^{2}\left(2\pi\nu\theta_{1}\right)+2\pi\nu\theta_{1}\sin\left(4\pi\nu\theta_{1}\right)+4\pi^{2}\nu^{2}\theta_{1}^{2}\cos^{2}\left(2\pi\nu\theta_{1}\right)\right]$. 

Because of the large value of $\nu\theta_{1}$ \cite{Debye1} the
last term in the brackets considerably exceeds the preceding ones,
and one can write with sufficient approximation: 

\begin{equation}
T=\frac{e^{2}C^{2}}{8m_{0}}\left[\theta_{1}\cos\left(2\pi\nu\theta_{1}\right)\right]^{2}\label{eq30}
\end{equation}
Furthermore according to (\ref{eq25}) 

\begin{equation}
U=\frac{e^{2}C^{2}}{8m_{0}}\left[\theta_{1}\sin\left(2\pi\nu\theta_{1}\right)\right]^{2}\label{eq31}
\end{equation}
The photoelectrically liberated electron therefore has the energy: 

\[
T+U=\frac{e^{2}C^{2}}{8m_{0}}\theta_{1}^{2}
\]
or as a result of (\ref{eq29}) 

\begin{equation}
T+U=nh\nu
\end{equation}

\marginpar{116}This is consistent with Einstein's law. \textbf{}\footnote{\textbf{ }A difference between the Sommerfeld theory and Einstein's
one is the absence of the work function on the former. See section
III of \cite{Pelogia}.} For the liberation of the electron should happen in reality at maximum

\makeatletter%
\long\def\@makefnmark{%
\hbox {{\normalfont \@thefnmark }}}%
\makeatother
\setcounter{tempfootnote}{\value{footnote}}
\setcounter{footnote}{2}
\renewcommand{\thefootnote}{\alph{footnote}}

$T$, where $T$ reaches an integral multiple of $h\nu$ (\footnotemark).

\textbf{}\footnotetext{\ It is noteworthy that (\ref{eq30}) strictly applies.}

\setcounter{footnote}{\value{tempfootnote}}
\renewcommand{\thefootnote}{\arabic{footnote}}
\makeatletter%
\long\def\@makefnmark{%
\hbox {{\textsuperscript \@thefnmark }}}%
\makeatother

In Sommerfeld's theory the value of the kinetic energy at the relevant
point of time is also an integral multiple of $h\nu$; after all,
this applies with approximation

\begin{equation}
T=U
\end{equation}

Should the free electron after Einstein's law fly with this kinetic
energy $T$, then the following question can be answered only with
difficulty by Sommerfeld: \textbf{}\footnote{\textbf{ }At the time of publication of this paper, Millikan was
conducting his experiments on the photoelectric effect \cite{Milikan1,Milikan2,Milikan3,Milikan4},
that confirmed Einstein's hypothesis as a non-resonance phenomenon
with linear energy-frequency relation for the emitted electrons. } Where does the potential energy $U$ remain? Herein I also see a
reason to prefer our theory over the latter one. 
\begin{verse}
Institute of Physics of Sendai University.
\end{verse}

\end{document}